\documentclass[aps,preprint,nofootinbib]{revtex4}
\usepackage{amsmath}
\usepackage{graphicx}
\usepackage{color}
\usepackage{slashed}
\usepackage{subfigure}
\usepackage{tabularx}
\usepackage{booktabs}
\usepackage{mathtools}
\usepackage{hyperref}
\usepackage{multirow}
\allowdisplaybreaks[4]

\begin{document}

\title{NLO corrections to $J/\psi+c+\bar c$ photoproduction\\[0.7cm]}

\author{\vspace{1cm} Qi-Ming Feng$^{1}$ and Cong-Feng Qiao$^{1,2}$\footnote[1]{qiaocf@ucas.ac.cn}}

\affiliation{\small {$^1$ School of Physics, University of Chinese Academy of
Sciences, Beijing 100049, China\\
$^2$ Key Laboratory of Vacuum Physics of CAS, Beijing 100049, China}}

\author{~\\~\\~\\}

\begin{abstract}
\vspace{0.5cm}
Based on the factorization framework of nonrelativistic quantum chromodynamics, we study the associated $J/\psi+c+\bar c$ photoproduction process at next-to-leading order in $\alpha_s$ and leading order in the velocity expansion.
The total cross section and differential cross section in $p_T^2$, $W$ and $z$ are presented.
The results indicate that the next-to-leading order corrections are substantial, and testable in experiment.

\end{abstract}
\maketitle

\newpage

\section{Introduction}

The study of heavy quarkonium system presents an exceptional opportunity to explore the nuances of the phenomena quantum chromodynamics (QCD) involved. Especially, charmonium provides a peculiar playground for the study of flavor physics and QCD in the charm sector, of which a huge amount of experimental data have been accumulated. The moderate charm energy enables the perturbative QCD(pQCD) calculation reliable to some extent, while poses a challenge to higher order pQCD corrections.

Nonrelativistic QCD (NRQCD)~\cite{Bodwin:1994jh} provides a consistent theoretical framework for the study of quarkonium production and decays. In NRQCD, the quarkonium production and decay processes are factorized into two sectors: the perturbative generation and decay of heavy quark pairs, the dominant quarkonium quark components, and the nonperturbative hadronization or dehadronization of these heavy quarks. The perturbative contributions are represented by the matching coefficients to pQCD calculation, namely the short-distance coefficients (SDCs), while the nonperturbative hadronization is described by the matrix elements of process-independent effective operators, known as the long-distance matrix elements (LDMEs).

Nevertheless, there are still some unsolved problems pending in the application and understanding of NRQCD. The NRQCD factorization formalism suggests that besides the leading Fock state contribution usually corresponding to the color singlet mechanism (CSM), higher Fock state contributions, such as the color octet mechanism (COM), emerge in the expansion of heavy quark relative velocity $v$ ($v\ll 1$).
The proposal of COM effectively reduces the discrepancies in $J/\psi$ production between next-to-leading order (NLO) CSM predictions and experimental results across $e^+e^-$ collisions at B factories, photoproduction at DESY HERA, and hadroproduction at Fermilab Tevatron and CERN LHC~\cite{Chang:2009uj,Artoisenet:2009xh,Campbell:2007ws,Gong:2008sn,Lansberg:2010vq,Kramer:1994zi,Kramer:1995nb,Butenschoen:2009zy,Butenschoen:2011ks,Butenschoen:2012px,Chao:2012iv,Ma:2010jj,Ma:2010yw,Zhang:2009ym,Gong:2012ug}.
However, the COM introduces considerable uncertainties. In Ref.~\cite{Bodwin:2012ft}, comparisons between two LDMEs fitted through different procedures in various collision processes show somehow incompatible results. Since different processes rely on distinct sets of LDME data, the process-independence of COM LDMEs is challenged. Some new methods for fitting COM LDMEs have been proposed later, but discussing them in detail is beyond the scope of this text and will be skipped here.

Experimental and theoretical inquiries into inclusive heavy quarkonium production have spanned several decades (see \cite{Brambilla:2010cs,Lansberg:2019adr,QuarkoniumWorkingGroup:2004kpm} for reviews).
Recent studies of $J/\psi$ photoproduction in electron-proton  ($ep$) collisions indicate that CS contributions, such as intrinsic charm~\cite{Flore:2020jau} and higher-order processes like $J/\psi + c + \bar{c}$~\cite{Li:2019nlr}, are evident.\@
Inspired by the fact that production processes akin to NLO QCD corrections exhibit notable contributions~\cite{Campbell:2007ws,Chen:2016hju,Yang:2022yxb}, we posit that the NLO contributions of the aforementioned 3-body final states photoproduction process $\gamma+g\to J/\psi + c + \bar{c}$ at $ep$ colliders remains relatively significant.
Since the $J/\psi+c+\bar c$ final state is experimentally detectable, theoretical analysis of HERA data holds significance and provides insights for future $ep$ colliders like EIC, EicC, and LHeC (FCC-eh).

In this work, based on the framework of NRQCD, we systematically compute the NLO corrections to the photoproduction process $\gamma + g \to J/\psi + c + \bar{c}$ at leading $v$ expansion in $ep$ collisions. The structure of this paper is organized as follows. In Section II, we detail the formalism and calculation of the concerned process. In Section III, the results of numerical evaluation are presented. The last section is reserved for the summary and conclusions.

\section{Formalism and Calculation}
Within the framework of NRQCD, the cross section for the photoproduction process at leading $v$ in $ep$ collision can be formulated as:
\begin{align}\label{cross_section}
    d\sigma (ep\to J/\psi+c+\bar c) = \int &dx d\eta f_{\gamma/e}(x,Q^2_{\max}) f_{g/p}(\eta,\mu^2)\nonumber\\&\!\!\times d\sigma (\gamma +g\to c\bar c[^3\!S_1]+c+\bar c)\langle \mathcal{O}^{J/\psi}(^3\!S_1) \rangle\ .
\end{align}
Here, $\langle \mathcal{O}^{J/\psi}(^3\!S_1) \rangle$ is the LDME of a $c\bar c[^3\!S_1]$ pair hadronizing into a $J/\psi$ meson.
$f_{g/p}(\eta,\mu^2)$ is the parton distribution function (PDF) of the incident gluon, and $\mu$ is the corresponding factorization scale.
$f_{\gamma/e}(x,Q^2_{\max})$ is the Weizsacker-Williams approximation (WWA) function of the photon distribution, defined as:
\begin{align}
    f_{\gamma/e}(x,Q^2_{\max}) = \frac{\alpha}{2\pi}\left[\frac{1+{(1-x)}^2}{x}\ln\frac{Q^2_{\max}}{Q^2_{\min}(x)}+2m_e^2x\left(\frac{1}{Q^2_{\max}} - \frac{1}{Q^2_{\min}(x)}\right)\right]\ ,
    \label{WWA}
\end{align}
where $m_e$ is the electron mass.
$Q^2_{\min}=m_e^2 x^2/(1-x)$ and $Q^2_{\max}$ represents the minimal and maximum virtuality of the incident photon, respectively.

In the calculation of the concerned process, the spinor helicity formalism~\cite{Kleiss:1985yh,Qiao:2003ue,Dixon:1996wi,Dixon:2013uaa,Arkani-Hamed:2017jhn} of the scattering amplitude and the conventional amplitude squaring approach are introduced in evaluating diagrams.
Specifically, most of the diagrams are calculated in helicity amplitudes, except for the Coulomb divergent part, where the conventional amplitude squaring approach is employed.

The dipole subtraction method~\cite{Catani:1996vz,Catani:2002hc} is adopted to counteract the infrared (IR) poles.
Therefore, the total cross section at NLO can be expressed as:
\begin{align}
    \sigma_{tot} &= \int_{3\text{-}\mathrm{body}}( d\sigma^{\mathrm{LO}} + d\sigma^{\mathrm{Virtual}} +d\sigma^{C} + \int_{1\text{-}\mathrm{body}} d\sigma^{A}) + \int_{4\text{-}\mathrm{body}} (d\sigma^{\mathrm{Real}} - d\sigma^{A})\ ,
    \label{sigma_parts}
\end{align}
where $d\sigma^{\mathrm{LO}}$, $d\sigma^{\mathrm{Virtual}}$, and $d\sigma^{\mathrm{Real}}$ are the LO, virtual, and real contributions to the cross section. $d\sigma^{C}$ represents the collinear subtraction counterterm arising from the redefination of the parton distributions. The contribution $d\sigma^{A}$ represents the dipole counterterm.
The correspondence of the various terms in (\ref{sigma_parts}) with helicity amplitudes goes as
\begin{align}
    &d\sigma^{\mathrm{LO}}\ ,\ d\sigma^{C}\ ,\ \int_{1\text{-}\mathrm{body}} d\sigma^{A}\propto \sum_{\alpha_1,\alpha_2,\alpha_3,\alpha_4,\alpha_5}|\mathcal{A}_{\mathrm{LO}}^{\alpha_1,\alpha_2,\alpha_3,\alpha_4,\alpha_5}|^2\ , \nonumber \\
    &d\sigma^{\mathrm{Virtual}}\propto \sum_{\alpha_1,\alpha_2,\alpha_3,\alpha_4,\alpha_5}2\mathrm{Re}\left[{(\mathcal{A}_{\mathrm{LO}}^{\alpha_1,\alpha_2,\alpha_3,\alpha_4,\alpha_5})}^* \mathcal{A}_{\mathrm{Virtual}}^{\alpha_1,\alpha_2,\alpha_3,\alpha_4,\alpha_5}\right]\ , \nonumber \\
    &d\sigma^{\mathrm{Real}}\propto \sum_{\alpha_1,\alpha_2,\alpha_3,\alpha_4,\alpha_5,\alpha_6}|\mathcal{A}_{\mathrm{Real}}^{\alpha_1,\alpha_2,\alpha_3,\alpha_4,\alpha_5,\alpha_6}|^2\ .
\end{align}
Here, $\alpha_{1,2,3,4,5,6}$ represent the helicities (polarizations) of on-shell particles, $\alpha_{1,2,4,5,6}\in{(+,-)}$ denote the helicities of initial photon, initial gluon, the final charm quark, the charm antiquark, and the final emitted gluon, $\alpha_{3}\in{(+,0,-)}$ signify the helicities of $J/\psi$ meson.
$\mathcal{A}_{\mathrm{LO}}^{\alpha_1,\alpha_2,\alpha_3,\alpha_4,\alpha_5}$, $\mathcal{A}_{\mathrm{Virtual}}^{\alpha_1,\alpha_2,\alpha_3,\alpha_4,\alpha_5}$, and $\mathcal{A}_{\mathrm{Real}}^{\alpha_1,\alpha_2,\alpha_3,\alpha_4,\alpha_5,\alpha_6}$ are helicity amplitudes of LO, virtual corrections, and real corrections, respectively.

\begin{figure}
    \includegraphics[width=13 cm]{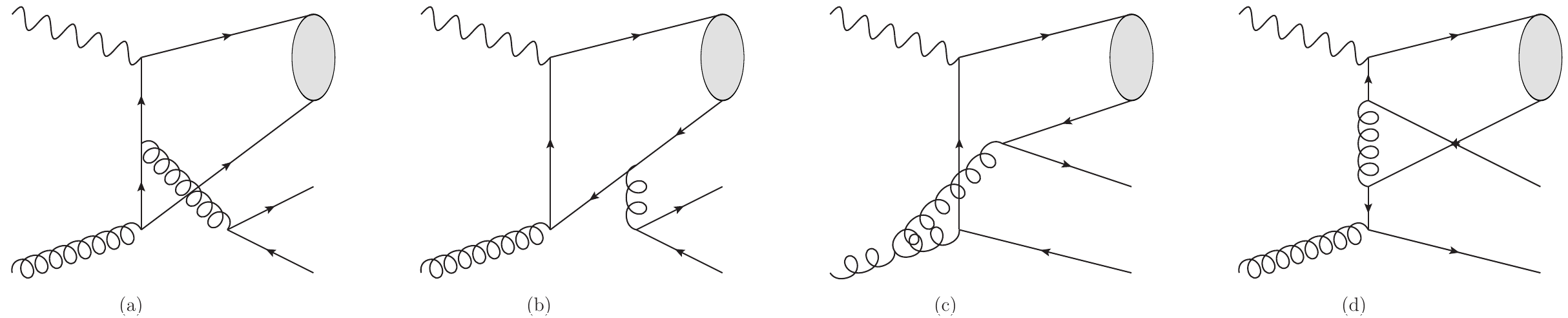}
    \caption{Typical LO Feynman diagrams for $\gamma+g\to J/\psi+c+\bar c$. All other diagrams can be generated by: 1.\ exchanging initial photon and gluon in $(a)$, $(b)$ and $(d)$; 2.\ reversing fermion lines; 3.\ constraining $J/\psi$ in other quark-antiquark pairs. There are 6 diagrams related to (a), 12 diagrams related to (b), 4 diagrams related to (c), and 8 diagrams related to (d). Diagrams with no contribution in the end are neglected. }\label{fig_Feynlo}
\end{figure}

There are 30 non-zero Feynman diagrams at the leading order of the concerned process, as schematically shown in FIG.~\ref{fig_Feynlo}.
Helicity amplitude of the LO contribution can be expressed as:
\begin{align}
    \mathcal{A}_{\mathrm{LO}}^{\alpha_1,\alpha_2,\alpha_3,\alpha_4,\alpha_5} = \sum_{i=1}^{30}\mathcal{A}^{\alpha_1,\alpha_2,\alpha_3,\alpha_4,\alpha_5}_i = \sum_{j=1}^{59}\mathcal{C}_j(p_a\cdot p_b, p_a\cdot \varepsilon^{\alpha_b}_b, \varepsilon^{\alpha_a}_a\cdot \varepsilon^{\alpha_b}_b)\mathcal{\hat A}_{j}^{\alpha_1,\alpha_2,\alpha_3,\alpha_4,\alpha_5}\ ,
    \label{LOampall}
\end{align}
where $p_a$ and $p_b$ represent on-shell momenta, $\varepsilon^{\alpha_a}_a$ and $\varepsilon^{\alpha_b}_b$ represent polarization vectors. The total helicity amplitude $\mathcal{A}_{\mathrm{LO}}^{\alpha_1,\alpha_2,\alpha_3,\alpha_4,\alpha_5}$ sums over all 30 LO diagrams $\mathcal{A}_i^{\alpha_1,\alpha_2,\alpha_3,\alpha_4,\alpha_5}$. We rearrange the summation into 59 distinct combinations of spinor products for simplification, expressed as $\mathcal{\hat A}_{j}^{\alpha_1,\alpha_2,\alpha_3,\alpha_4,\alpha_5}$. $\mathcal{C}_j$ represents corresponding coefficients of $\mathcal{\hat A}_j^{\alpha_1,\alpha_2,\alpha_3,\alpha_4,\alpha_5}$, composed of scalar products of momenta and polarization vectors.

\begin{figure}[tbh]
    \includegraphics[width=0.6\textwidth]{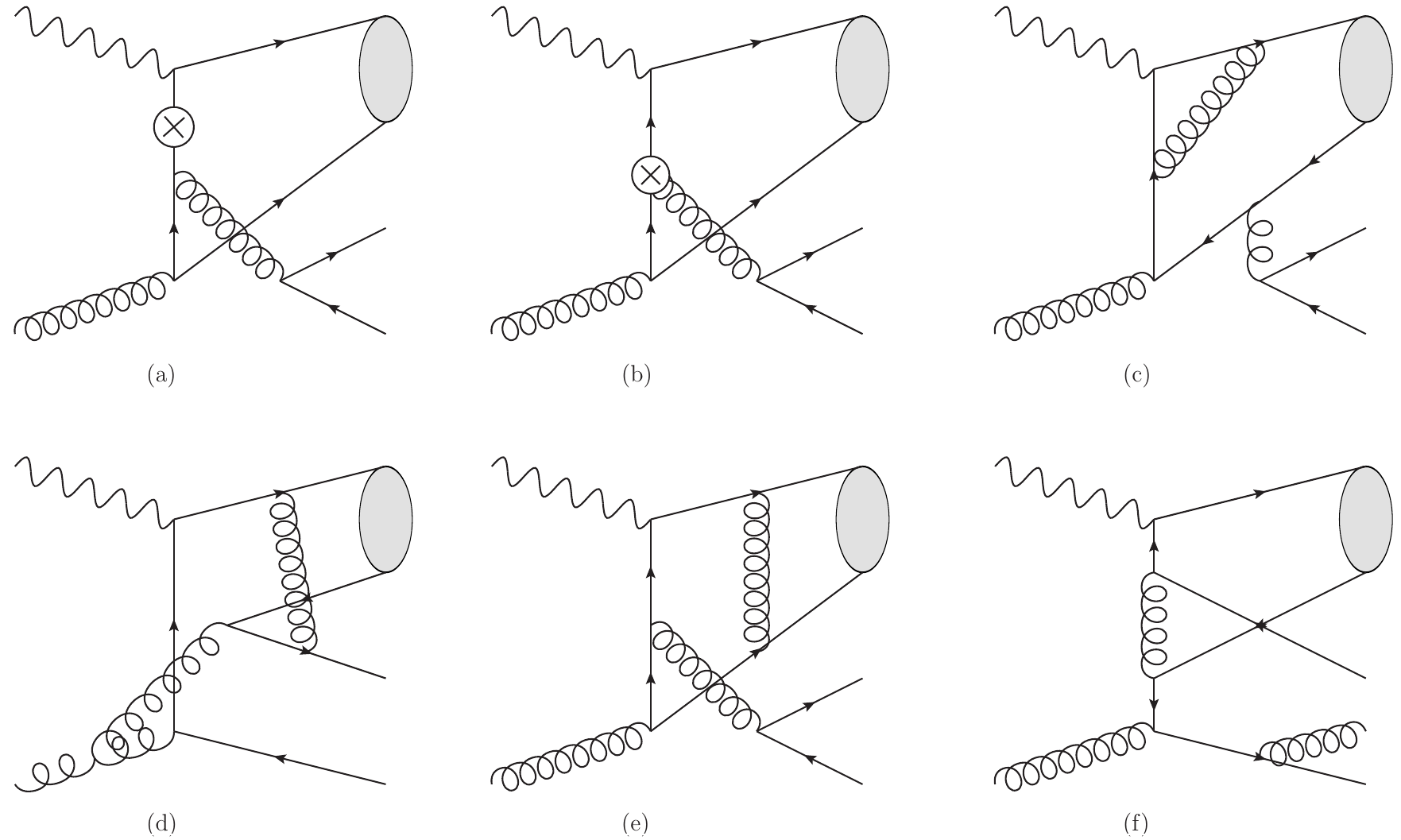}
    \caption{Typical NLO Feynman diagrams of $\gamma+g\to J/\psi+c+\bar c$. $(a)$ and $(b)$ are typical renormalizaiton diagrams, where the $\otimes$ denotes the counterterm of a propagator or a vertex. $(c)\sim(e)$ are typical loop diagrams with UV, IR and Coulomb poles, respectively. The UV pole in (c) can be canceled by introducing renormalization. The IR pole in $(d)$ can be counteracted with real corrections (initial gluon introduced pole) and other loop diagrams (the pole introduced form $J/\psi$ and $c$). The coulomb pole in $(e)$ can be factorization out. And $(f)$ is a typical diagram of the real correction process $\gamma+g\to J/\psi+c+\bar c+g$. }\label{fig_nlo}
\end{figure}

In the calculation of NLO corrections, the 't Hooft-Veltman scheme (HV) dimensional regularization (DR) is employed.
Some of the NLO diagrams are shown in FIG.~\ref{fig_nlo}.
Loop diagrams are evaluated using two distinct methods, the integrand reduction and IBP reduction.
Helicity amplitudes of loop diagrams without Coulomb divergence are evaluated using integrand reduction via Laurent-expansion method~\cite{Mastrolia:2012bu} by the semi-numerical reduction C++ library Ninja~\cite{Peraro:2014cba}.
For the Coulomb divergent diagrams, the Coulomb term can be expressed as:
\begin{eqnarray}\label{DPs}
    \frac{1}{[{(-\frac{p_3}{2}+q)}^2-m_c^2]q^2[(\frac{p_3}{2}+q)^2-m_c^2]}
    %&\nonumber\\ &\!\!\!\!\!\!\!\!\!\!\!\!\!\!\!\!\!\!\!\!\!\!\!\! 
    = \frac{1}{2}\left(\frac{1}{{(q^2)}^2[(-\frac{p_3}{2}+q)^2-m_c^2]} +
    \frac{1}{{(q^2)}^2[{(\frac{p_3}{2}+q)}^2-m_c^2]}\right) ,~~
\end{eqnarray}
where only the loop propagators related to the Coulomb poles are depicted. $p_3$ is $J/\psi$ meson momentum, $q$ is the loop momentum, and $m_c$ is the charm quark mass.
The introduction of exceptional higher-power loop propagators $\frac{1}{{(q^2)}^2}$ prevents the evaluation of Coulomb divergent diagrams using Ninja.
Therefore, we employ the integration by parts (IBP) reduction method to evaluate Coulomb loops by NeatIBP~\cite{Wu:2023upw}.

Ultraviolet (UV) singularities are canceled by renormalization.
In our calculation, renormalizaiton constant of the QCD coupling constant $Z_g$ is defined in the modified minimal subtraction ($\overline{\rm MS}$) scheme, renormalizaiton constant of the charm quark field $Z_2$ and mass $Z_m$ and the gluon field $Z_3$ are defined in the on-shell (OS) schem.
The counterterms are given by:
\begin{align}
&\delta Z_2^{\rm OS}=-C_F\frac{\alpha_s}{4\pi}\left[\frac{1}{\epsilon_{\rm {UV}}}+\frac{2}{\epsilon_{\rm {IR}}}-3\gamma_E+3\ln\frac{4\pi\mu_r^2}{m^2}+4\right]\ ,\nonumber\\
&\delta Z_3^{\rm OS}=\frac{\alpha_s}{4\pi}\left[(\beta'_0-2C_A)(\frac{1}{\epsilon_{\rm {UV}}}-\frac{1}{\epsilon_{\rm {IR}}})-\frac{4}{3}T_f\left(\frac{1}{\epsilon_{\rm{UV}}}-\gamma_E+\ln \frac{4\pi\mu_r^2}{m^2}\right)\right]\ ,\nonumber\\
&\delta Z_m^{\rm OS}=-3C_F\frac{\alpha_s}{4\pi}\left[\frac{1}{\epsilon_{\rm UV}}-\gamma_E+\ln\frac{4\pi\mu_r^2}{m^2} +\frac{4}{3}\right]\ , \nonumber\\
&\delta Z_g^{\overline{\rm MS}}=-\frac{\beta_0}{2}\frac{\alpha_s}{4\pi}\left[\frac{1}{\epsilon_{\rm UV}} -\gamma_E + \ln(4\pi)\right]\ .
\label{cts}
\end{align}
Here, $\beta'_0=(11/3)C_A-(4/3)T_f n'_f$ is the one-loop coefficient of the QCD beta function, $n'_f=3$ is the number of light quarks, $\gamma_E$ is Euler's constant, $m$ represents the mass of charm quark, $C_A$ and $T_f$ attribute to the color $SU(3)$ group, $n_f=4$ is the number of active quarks, $\mu_r$ denotes the renormalization scale.

In the evaluation of real corrections, as expected, the dipole counterterm $d\sigma^{A}$ exhibits the same pointwise singular behaviour as $d\sigma^{\mathrm{Real}}$, and $\int_{1\text{-}\mathrm{body}}d\sigma^{A}$ stands for the analytic integration of $d\sigma^{A}$ over the phase space with an additional real gluon in dimension $D=4-2\varepsilon$, cancelling out the remaining analytic $\frac{1}{\varepsilon}$ and $\frac{1}{\varepsilon^2}$ divergences in virtual correction.

In the concerned process, the dipole terms associated with quarkonium cancel out.
As a result, only 3 types of dipole terms remain:
\begin{enumerate}
    \item initial gluon emitter with final charm (anti-charm) spectator: $\mathcal{D}^{gg}_{c},\ \mathcal{D}^{gg}_{\bar c}$,
    \item final charm (anti-charm) emitter with initial gluon spectator: $\mathcal{D}^{g}_{cg},\ \mathcal{D}^{g}_{\bar cg}$,
    \item final charm (anti-charm) emitter with final anti-charm (charm) spectator: $\mathcal{D}_{cg,\bar c},\ \mathcal{D}_{\bar c g,c}$,
\end{enumerate}
Here, dipole contributions $\mathcal{D}^{gg}_{c}$, $\mathcal{D}^{gg}_{\bar c}$, $\mathcal{D}^{g}_{cg}$, $\mathcal{D}^{g}_{\bar cg}$, $\mathcal{D}_{cg,\bar c}$, and $\mathcal{D}_{\bar c g,c}$ are defined in Ref.~\cite{Catani:2002hc}.
Hence, the dipole factorization form of $d\sigma^A$ writes:
\begin{align}
    d\sigma^A=d\Gamma^{(4)}\left(\sum_{\substack{i,k=c,\bar c\\i\neq k}}\mathcal{D}_{ig,k} + \sum_{i=c,\bar c}\mathcal{D}_{ig}^g + \sum_{k=c,\bar c}\mathcal{D}_{k}^{gg}\right)
\end{align}
with $d\Gamma^{(4)}$ being the 4-body phase space including all factors of QCD independent.
The integrated dipoles are obtained from $(5.23)$, $(5.56)$, and $(5.88)$ of Ref.~\cite{Catani:2002hc}.

Of the cancellation of divergences, we extract the IR divergences in loop diagrams by means of the method developed in Ref.~\cite{Dittmaier:2003bc}, which ensures all IR divergences in one-loop diagrams with more than three loop propagators being expressed as sum of triangles (diagrams with three loop propagators).
For example, for an $N$-point loop ($N\geq 3$)
\begin{align}\label{divseprat}
    \int_{\rm{div}} d^{4-2\epsilon}q\prod_{i=0}^{N-1}\frac{1}{D_i} = \int_{\rm{div}} d^{4-2\epsilon}q \sum_{i=0}^{N-1}\sum_{\substack{j=0\\k\neq i,i+1}}^{N-1}\frac{A_{ij}}{D_i D_{i+1} D_j}\ ,
\end{align}
where $D_i$ represents loop propagators. $A_{ij}$ represents the corresponding coefficients.
In the calculation, divergences in virtual correction and dipole contribution are analytically canceled out.

\section{Results}
In our numerical calculation, the charm mass takes half of $J/\psi$ mass, i.e. $m_c=1.5\ \rm{GeV}$.
The renormalization scale $\mu_r$ and the factorizaiton scale $\mu_f$ are set as $\mu_r=\mu_f=m_{T}\equiv\sqrt{p_{T}^2+4m_{c}^2}$, where $m_T$ is the $J/\psi$ transverse mass.
Theoretical uncertainties are estimated by varying the charm quark mass in $m_c = 1.5 \pm 0.1\ \rm{GeV}$ and the scales in the interval $\frac{1}{2}m_{T} \leq \mu_r, \mu_f \leq 2m_{T}$.
The running coupling constant is determined by the one-loop (two-loop) formula at LO (NLO), and the PDF set CTEQ6L1 (CTEQ6M)~\cite{Pumplin:2002vw} is used at LO (NLO).
The LDME follows $\langle \mathcal{O}^{J/\psi}(^3\!S_1) \rangle = 2(2J+1)N_c|R(0)|^2/{4\pi}$ with $J=1$ for the $J/\psi$ meson, and $N_c=3$ is the number of color charges.
The radial wave function $|R(0)|^2=1.01\ \rm{GeV}^3$ is extract form $\Gamma(J/\psi\to e^+e^-)=5.55\ \rm{keV}$~\cite{ParticleDataGroup:2020ssz}, thus, we have $\langle \mathcal{O}^{J/\psi}(^3\!S_1) \rangle=1.45\ \rm{GeV}^3$.

The collision energy is set accroding to the HERA collider: $27.5\ \rm{GeV}$ for eletrons (positrons) and $920\ \rm{GeV}$ for protons.
The photon virtuality is constrainted to $Q^2_{\max}\leq2.5\ \rm{GeV}^2$.
To exclude resolved photoproduction and diffractive production of $J/\psi$, experimental cuts based on the H1 collaboration measurement~\cite{H1:2010udv} are applied: $p_T>1\ \rm{GeV}$, $60\ \rm{GeV}<W<240\ \rm{GeV}$, and $0.3<z<0.9$. Here, $W=\sqrt{{(p_\gamma+p_{p})}^2}$ is the mass of the hadronic final state, $z=(p_{3}\cdot p_{p})/(p_{\gamma}\cdot p_{p})$ is the elasticity of the $J/\psi$ meson production process, and $p_{p}$, $p_{\gamma}$ are the momenta of the incident proton and photon, respectively.
Additionally, the feed-down contribution from the $\psi'$ is taken into account, which yields an enhancing factor about 0.278.

\begin{table}
    \caption{Scale and mass dependence of the total cross section at LO (expressed in $\rm{nb}$) in various PDF sets without feed-down contribution. Here, $\mu=\mu_r=\mu_f$. }
    \centering
    {\small
    \begin{tabular}{p{3cm}<{\centering} || p{2.5cm}<{\centering} | p{2cm}<{\centering} p{2cm}<{\centering} p{2cm}<{\centering}}
    \toprule[2pt]
      PDF sets  & $m_c\backslash \mu$ & $\frac{1}{2}m_T$ & $m_T$ & $2m_T$  \\
    \toprule[1pt]
    \multirow{3}{*}{CTEQ6L1}    & $1.4\ \rm{GeV}$ & $0.185$ & $0.111$ & $0.070$  \\
                                & $1.5\ \rm{GeV}$ & $0.123$ & $0.074$ & $0.046$  \\
                                & $1.6\ \rm{GeV}$ & $0.084$ & $0.049$ & $0.031$  \\
                                \bottomrule[1pt]
    \multirow{3}{*}{CT14LO}     & $1.4\ \rm{GeV}$ & $0.128$ & $0.073$ & $0.046$  \\
                                & $1.5\ \rm{GeV}$ & $0.084$ & $0.048$ & $0.030$  \\
                                & $1.6\ \rm{GeV}$ & $0.056$ & $0.032$ & $0.020$  \\
                                \bottomrule[1pt]
    \multirow{3}{*}{CTEQ6M}     & $1.4\ \rm{GeV}$ & $0.109$ & $0.066$ & $0.043$  \\
                                & $1.5\ \rm{GeV}$ & $0.073$ & $0.044$ & $0.029$  \\
                                & $1.6\ \rm{GeV}$ & $0.051$ & $0.030$ & $0.019$  \\
                                \bottomrule[1pt]
    \multirow{3}{*}{CT14NLO}    & $1.4\ \rm{GeV}$ & $0.095$ & $0.063$ & $0.042$  \\
                                & $1.5\ \rm{GeV}$ & $0.066$ & $0.042$ & $0.028$  \\
                                & $1.6\ \rm{GeV}$ & $0.045$ & $0.029$ & $0.019$  \\
                                \bottomrule[1pt]
    \multirow{3}{*}{CT18NLO}    & $1.4\ \rm{GeV}$ & $0.094$ & $0.063$ & $0.042$  \\
                                & $1.5\ \rm{GeV}$ & $0.065$ & $0.042$ & $0.028$  \\
                                & $1.6\ \rm{GeV}$ & $0.045$ & $0.029$ & $0.019$  \\
                                \bottomrule[1pt]
    \end{tabular}
    }\label{tab_tcs}
\end{table}

As a result, the total cross section of the concerned process at NLO (LO) is
\begin{align}
\sigma_{tot}=0.118^{+0.168}_{-0.065}\ (0.074^{+0.111}_{-0.043})\ \rm{nb}.
\end{align}
Our LO result agrees with what in Ref.~\cite{Li:2009zzu} after taking the same inputs.
NLO corrections yield a $K$ factor about $1.60$, which is a prominent enhancement of the cross section.
It is evident that the error on the total cross section is large.
In Table~\ref{tab_tcs}, we present the errors of the LO total cross section on the charm quark mass and scales across various PDF sets.
The cross section with each PDF set shows strong sensitivity to both charm quark mass and scales, particularly to scales. Results using the NLO PDF sets CTEQ6M, CT14NLO~\cite{Dulat:2015mca}, and CT18NLO~\cite{Hou:2019qau} exhibit stronger dependence on scales compared to those obtained using LO PDF sets CTEQ6L1 and CT14LO~\cite{Dulat:2015mca}.

In estimating the number of the concerned process events at HERA, we consider that the H1 collaboration reconstructed the $J/\psi$ meson candidates through the decay channel $J/\psi\to\mu^+\mu^-$ in the photoproduction process, and the photoproduction sample corresponds to an integrated luminosity of $\mathcal{L}=165\ \rm{pb}^{-1}$~\cite{H1:2010udv}.
Accroding to our numerical results, with a branching fraction of $\Gamma(J/\psi\to\mu^+\mu^-)/\Gamma_{tot}\simeq6\%$, the number of reconstructed $J/\psi+c+\bar c$ events in the photoproduction process at NLO (LO) is about $521\sim 2833\ (304\sim 1829)$ at HERA.\@
Here, we omit the unclear tagging efficiency of $c$-jets.

\begin{figure}
    \centering
    \subfigure[]{
    \includegraphics[width=0.48\textwidth]{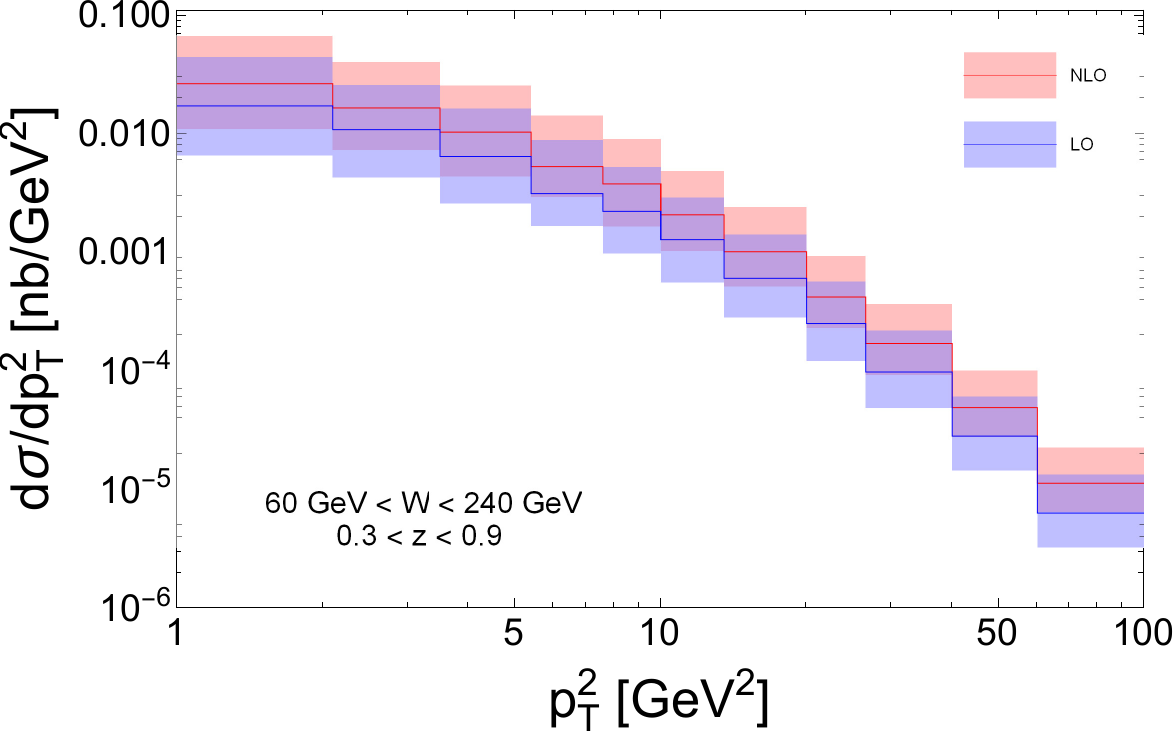}}
    \subfigure[]{
    \includegraphics[width=0.48\textwidth]{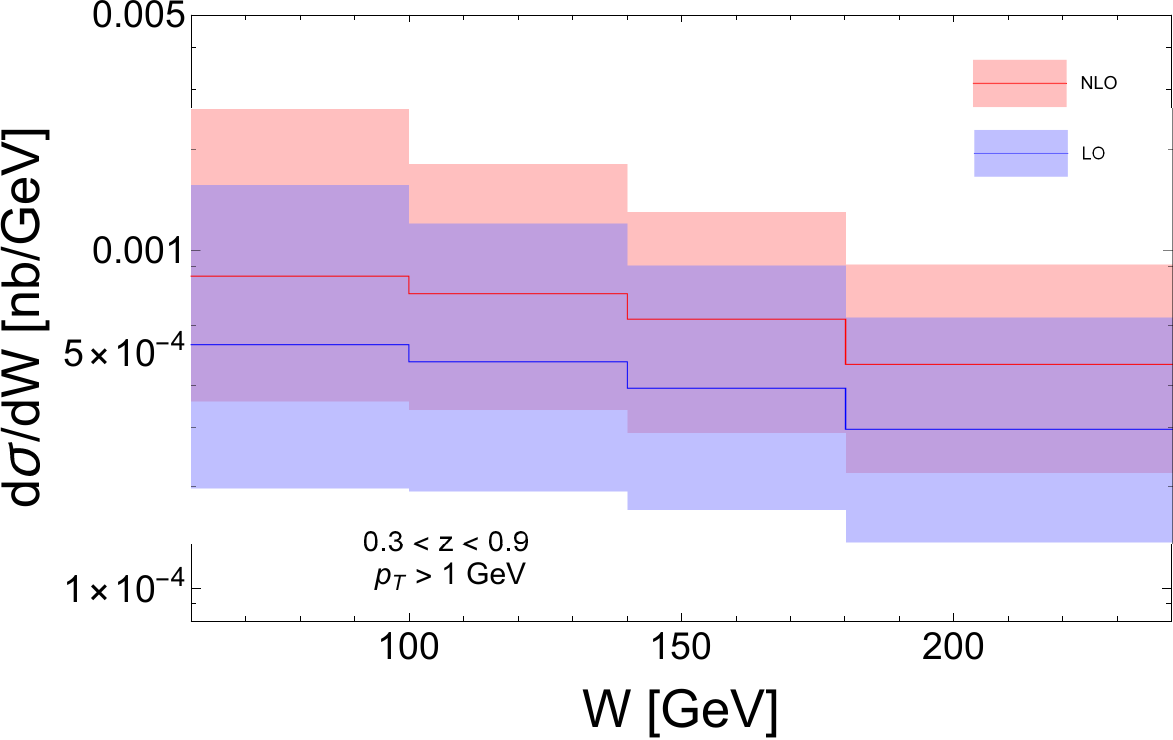}}
    \subfigure[]{
    \includegraphics[width=0.48\textwidth]{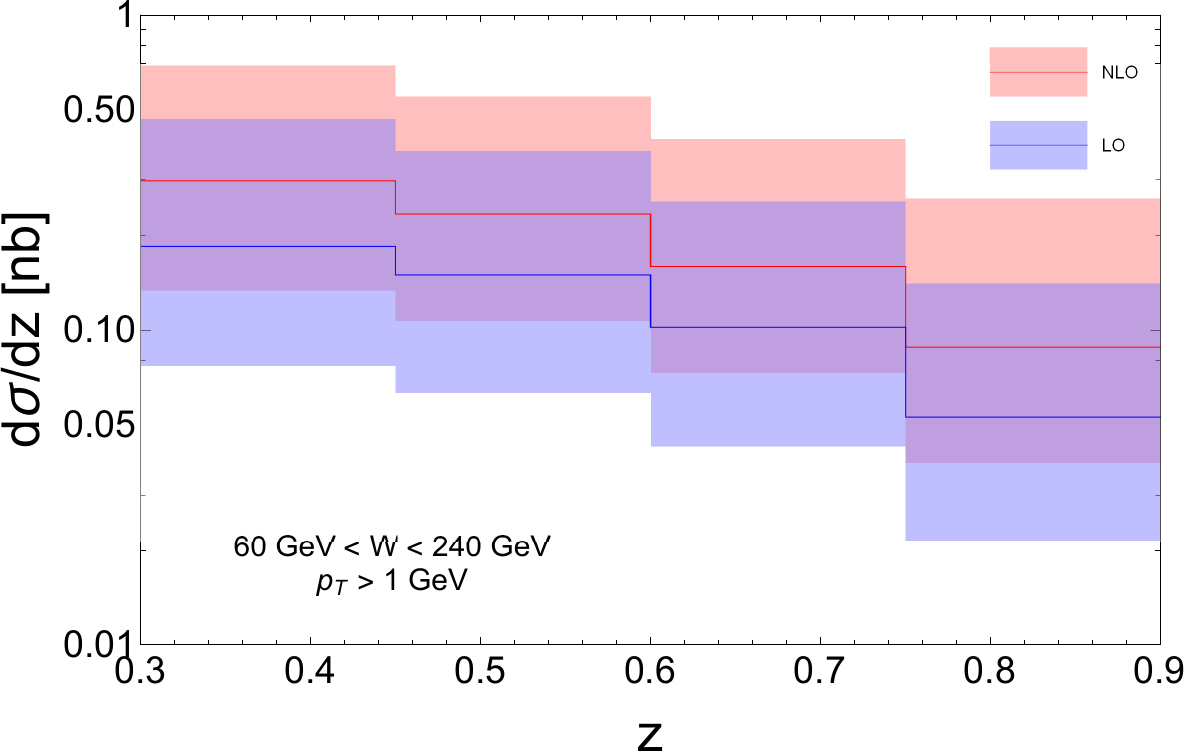}}
    \caption{The differential cross section in (a) $p_{T}^2$, (b) $W$, and (c) $z$ distributions of the photoproduction process $\gamma+g\to J/\psi+c+\bar c$ at LO and NLO in the CSM.\@ The shaded bands indicate the theoretical uncertainties with upper bound for $\mu_r,\mu_f=\frac{1}{2}m_T$ and $m_c=1.4\ \rm{GeV}$ and lower bound for $\mu_r,\mu_f=2m_T$ and $m_c=1.6\ \rm{GeV}$. }\label{dists}
\end{figure}

The differential cross section distributions in $p_T^2$, $W$, and $z$, are presented in FIG.~\ref{dists} $(a)\sim (c)$, respectively.
The $p_T^2$ distribution of $J/\psi$, as shown in FIG.~\ref{dists}$(a)$, is presented in the range $1\ \rm{GeV}^2<p_T^2<100\ \rm{GeV}^2$.
Compared to the $p_T^2$ distribution of the inclusive process $\gamma+g\to J/\psi+X$ at NLO CSM in Ref.~\cite{Butenschoen:2009zy}, the $J/\psi+c+\bar c$ process have a much milder drop as $p_T^2$ increases.
Although the differential cross section of the concerned process is much lower than that of the $J/\psi+X$ production process at low-$p_T$, in the region of $60\ \rm{GeV} < p_T^2 < 100\ \rm{GeV}^2$, the ratio $\sigma(\gamma+g\to J/\psi+c+\bar c)/\sigma(\gamma+g\to J/\psi+X)$ tends to about $1/2$.
It means that the concerned process is a not negligible contribution in the evaluation of $J/\psi$ photoproduction processes in $ep$ colliders, particularly in the large-$p_T$ region.

The $W$ and $z$ distributions in FIG.~\ref{dists}$(b)\sim(c)$ significantly undershoot the inclusive $\gamma+g\to J/\psi+X$ process in the CSM in Ref.~\cite{Butenschoen:2009zy}.
The $W$ distribution of the concerned process, the CS $J/\psi+X$ production process, and the H1 measurement show a similar trend.
However, the $z$ distribution shows a completely different trend compared to the CS $J/\psi+X$ production process and H1 data.
In contrast to the moderate $z$ distribution trends in Ref.\cite{Butenschoen:2009zy} and Ref.\cite{H1:2010udv}, the concerned process exhibits a rapid decrease in distribution with increasing $z$, particularly at large $z$.

The discrepancy on $z$ distribution may originate from the distinct dynamics of these processes. With the elasticity $z$ increases, the $J/\psi$ meson tends to parallel with the incident photon, that makes the momentum of the sum of the final charm quark and final anti-charm quark reduced.
Since the mass of the sum momentum greater than the sum of charm quark mass and anti-charm quark mass: $m_{c+\bar c}=\sqrt{{(p_c+p_{\bar c})}^2}>2m_c$, the dynamics of the two (anti-) charm quarks are constrainted, apparently, the cross section is also limited.
For the $\gamma+g\to J/\psi+X$ process, $X=g,q$ is a light particle, there is no such constraint in $z$, so the $z$-distribution is flat.
This can be confirmed by comparing upper, middle, and lower bounds.
As $m_c$ drops from $1.6\ \rm{GeV}$ to $1.4\ \rm{GeV}$, the distribution tends to be more flat.

\section{Summary}
In this work, we investigate the photoproduction of $J/\psi+c+\bar c$ in $ep$ collider HERA at NLO QCD and LO in $v$ within the framework of NRQCD.\@
Numerical result shows that the NLO QCD corrections are prominent in the concerned process.
We also present the differential cross section in $p_T$, $W$ and $z$.
The concerned process shows a significantly contribution in $J/\psi$ photoproduction processes at large-$p_T$.\@
Comparing with the $p_T$ distribution of the inclusive $\gamma+g\to J/\psi+X$ process in the CSM, the concerned process have a much milder drop as $p_T$ increases.
The trend in the $W$ distribution typically agrees with that of the CS $J/\psi+X$ photoproduction process and H1 data, by a factor of approximately $1/10$.
The $z$ distribution shows a severe drop, which differs from that of the inclusive process and H1 data.
This difference may originate from the distinct dynamics of these processes, and expects a further data analysis.
Future $ep$ colliders, such as EIC, EicC, and LHeC (FCC-eh), are expected to have rich production of the concerned process due to their high luminosity. Furthermore, some of them are capable of detecting particle polarizations, theoretical analysis on the polarizations of the concerned process awaits future investigation.

%%%%%%%%%%%%%%%%%%%%%%%%%%%%%%%%%%%%%%%%%%%%%%%%%%%%%%%%%%%%%%%%%%%%%
\vspace{1.0cm} {\bf Acknowledgments}

This work was supported in part by the National Key Research and Development Program of China under Contracts Nos.2020YFA0406400,
and the National Natural Science Foundation of China (NSFC) under the Grant 12235008.
%%%%%%%%%%%%%%%%%%%%%%%%%%%%%%%%%%%%%%%%%%%%%%%%%%%%%%%%%%%%%%%%%%%%


\begin{thebibliography}{9}

%\cite{Bodwin:1994jh}
\bibitem{Bodwin:1994jh}
G.~T.~Bodwin, E.~Braaten and G.~P.~Lepage,
%``Rigorous QCD analysis of inclusive annihilation and production of heavy quarkonium,''
Phys. Rev. D \textbf{51} (1995), 1125-1171
[erratum: Phys. Rev. D \textbf{55} (1997), 5853]
doi:10.1103/PhysRevD.55.5853
[arXiv:hep-ph/9407339 [hep-ph]].
%2648 citations counted in INSPIRE as of 09 Apr 2022

%\cite{Chang:2009uj}
\bibitem{Chang:2009uj}
C.~H.~Chang, R.~Li and J.~X.~Wang,
%``J/psi polarization in photo-production up-to the next-to-leading order of QCD,''
Phys. Rev. D \textbf{80}, 034020 (2009)
doi:10.1103/PhysRevD.80.034020
[arXiv:0901.4749 [hep-ph]].
%48 citations counted in INSPIRE as of 09 Mar 2024

%\cite{Artoisenet:2009xh}
\bibitem{Artoisenet:2009xh}
P.~Artoisenet, J.~M.~Campbell, F.~Maltoni and F.~Tramontano,
%``J/psi production at HERA,''
Phys. Rev. Lett. \textbf{102}, 142001 (2009)
doi:10.1103/PhysRevLett.102.142001
[arXiv:0901.4352 [hep-ph]].
%68 citations counted in INSPIRE as of 09 Mar 2024

%\cite{Campbell:2007ws}
\bibitem{Campbell:2007ws}
J.~M.~Campbell, F.~Maltoni and F.~Tramontano,
%``QCD corrections to J/psi and Upsilon production at hadron colliders,''
Phys. Rev. Lett. \textbf{98}, 252002 (2007)
doi:10.1103/PhysRevLett.98.252002
[arXiv:hep-ph/0703113 [hep-ph]].
%261 citations counted in INSPIRE as of 09 Mar 2024

%\cite{Gong:2008sn}
\bibitem{Gong:2008sn}
B.~Gong and J.~X.~Wang,
%``Next-to-leading-order QCD corrections to $J/\psi$ polarization at Tevatron and Large-Hadron-Collider energies,''
Phys. Rev. Lett. \textbf{100}, 232001 (2008)
doi:10.1103/PhysRevLett.100.232001
[arXiv:0802.3727 [hep-ph]].
%174 citations counted in INSPIRE as of 09 Mar 2024

%\cite{Lansberg:2010vq}
\bibitem{Lansberg:2010vq}
J.~P.~Lansberg,
%``QCD corrections to J/psi polarisation in pp collisions at RHIC,''
Phys. Lett. B \textbf{695}, 149-156 (2011)
doi:10.1016/j.physletb.2010.10.054
[arXiv:1003.4319 [hep-ph]].
%68 citations counted in INSPIRE as of 09 Mar 2024

%\cite{Kramer:1994zi}
\bibitem{Kramer:1994zi}
M.~Kramer, J.~Zunft, J.~Steegborn and P.~M.~Zerwas,
%``Inelastic J / psi photoproduction,''
Phys. Lett. B \textbf{348}, 657-664 (1995)
doi:10.1016/0370-2693(95)00155-E
[arXiv:hep-ph/9411372 [hep-ph]].
%131 citations counted in INSPIRE as of 09 Mar 2024

%\cite{Kramer:1995nb}
\bibitem{Kramer:1995nb}
M.~Kr\"amer,
%``QCD corrections to inelastic J / psi photoproduction,''
Nucl. Phys. B \textbf{459}, 3-50 (1996)
doi:10.1016/0550-3213(95)00568-4
[arXiv:hep-ph/9508409 [hep-ph]].
%238 citations counted in INSPIRE as of 09 Mar 2024

%\cite{Butenschoen:2009zy}
\bibitem{Butenschoen:2009zy}
M.~Butenschoen and B.~A.~Kniehl,
%``Complete next-to-leading-order corrections to J/psi photoproduction in nonrelativistic quantum chromodynamics,''
Phys. Rev. Lett. \textbf{104}, 072001 (2010)
doi:10.1103/PhysRevLett.104.072001
[arXiv:0909.2798 [hep-ph]].
%104 citations counted in INSPIRE as of 15 Mar 2024

%\cite{Butenschoen:2012px}
\bibitem{Butenschoen:2012px}
M.~Butenschoen and B.~A.~Kniehl,
%``J/psi polarization at Tevatron and LHC: Nonrelativistic-QCD factorization at the crossroads,''
Phys. Rev. Lett. \textbf{108}, 172002 (2012)
doi:10.1103/PhysRevLett.108.172002
[arXiv:1201.1872 [hep-ph]].
%235 citations counted in INSPIRE as of 09 Mar 2024

%\cite{Butenschoen:2011ks}
\bibitem{Butenschoen:2011ks}
M.~Butenschoen and B.~A.~Kniehl,
%``Probing nonrelativistic QCD factorization in polarized $J/\psi$ photoproduction at next-to-leading order,''
Phys. Rev. Lett. \textbf{107}, 232001 (2011)
doi:10.1103/PhysRevLett.107.232001
[arXiv:1109.1476 [hep-ph]].
%58 citations counted in INSPIRE as of 09 Mar 2024

%\cite{Chao:2012iv}
\bibitem{Chao:2012iv}
K.~T.~Chao, Y.~Q.~Ma, H.~S.~Shao, K.~Wang and Y.~J.~Zhang,
%``$J/\psi$ Polarization at Hadron Colliders in Nonrelativistic QCD,''
Phys. Rev. Lett. \textbf{108}, 242004 (2012)
doi:10.1103/PhysRevLett.108.242004
[arXiv:1201.2675 [hep-ph]].
%289 citations counted in INSPIRE as of 09 Mar 2024

%\cite{Ma:2010jj}
\bibitem{Ma:2010jj}
Y.~Q.~Ma, K.~Wang and K.~T.~Chao,
%``A complete NLO calculation of the $J/\psi$ and $\psi'$ production at hadron colliders,''
Phys. Rev. D \textbf{84}, 114001 (2011)
doi:10.1103/PhysRevD.84.114001
[arXiv:1012.1030 [hep-ph]].
%123 citations counted in INSPIRE as of 09 Mar 2024

%\cite{Ma:2010yw}
\bibitem{Ma:2010yw}
Y.~Q.~Ma, K.~Wang and K.~T.~Chao,
%``$J/\psi (\psi^\prime)$ production at the Tevatron and LHC at ${\cal O}(\alpha_s^4v^4)$ in nonrelativistic QCD,''
Phys. Rev. Lett. \textbf{106}, 042002 (2011)
doi:10.1103/PhysRevLett.106.042002
[arXiv:1009.3655 [hep-ph]].
%288 citations counted in INSPIRE as of 09 Mar 2024

%\cite{Zhang:2009ym}
\bibitem{Zhang:2009ym}
Y.~J.~Zhang, Y.~Q.~Ma, K.~Wang and K.~T.~Chao,
%``QCD radiative correction to color-octet $J/\psi$ inclusive production at B Factories,''
Phys. Rev. D \textbf{81}, 034015 (2010)
doi:10.1103/PhysRevD.81.034015
[arXiv:0911.2166 [hep-ph]].
%102 citations counted in INSPIRE as of 09 Mar 2024

%\cite{Gong:2012ug}
\bibitem{Gong:2012ug}
B.~Gong, L.~P.~Wan, J.~X.~Wang and H.~F.~Zhang,
%``Polarization for Prompt J/\ensuremath{\psi} and \ensuremath{\psi}(2s) Production at the Tevatron and LHC,''
Phys. Rev. Lett. \textbf{110}, no.4, 042002 (2013)
doi:10.1103/PhysRevLett.110.042002
[arXiv:1205.6682 [hep-ph]].
%233 citations counted in INSPIRE as of 09 Mar 2024

%\cite{Bodwin:2012ft}
\bibitem{Bodwin:2012ft}
G.~T.~Bodwin,
%``Theory of Charmonium Production,''
[arXiv:1208.5506 [hep-ph]].
%13 citations counted in INSPIRE as of 09 Mar 2024

%\cite{Brambilla:2010cs}
\bibitem{Brambilla:2010cs}
N.~Brambilla, S.~Eidelman, B.~K.~Heltsley, R.~Vogt, G.~T.~Bodwin, E.~Eichten, A.~D.~Frawley, A.~B.~Meyer, R.~E.~Mitchell and V.~Papadimitriou, \textit{et al.}
%``Heavy Quarkonium: Progress, Puzzles, and Opportunities,''
Eur. Phys. J. C \textbf{71}, 1534 (2011)
doi:10.1140/epjc/s10052-010-1534-9
[arXiv:1010.5827 [hep-ph]].
%1841 citations counted in INSPIRE as of 09 Mar 2024

%\cite{Lansberg:2019adr}
\bibitem{Lansberg:2019adr}
J.~P.~Lansberg,
%``New Observables in Inclusive Production of Quarkonia,''
Phys. Rept. \textbf{889}, 1-106 (2020)
doi:10.1016/j.physrep.2020.08.007
[arXiv:1903.09185 [hep-ph]].
%148 citations counted in INSPIRE as of 09 Mar 2024

%\cite{QuarkoniumWorkingGroup:2004kpm}
\bibitem{QuarkoniumWorkingGroup:2004kpm}
N.~Brambilla \textit{et al.} [Quarkonium Working Group],
%``Heavy quarkonium physics,''
doi:10.5170/CERN-2005-005
[arXiv:hep-ph/0412158 [hep-ph]].
%1023 citations counted in INSPIRE as of 09 Mar 2024

%\cite{Flore:2020jau}
\bibitem{Flore:2020jau}
C.~Flore, J.~P.~Lansberg, H.~S.~Shao and Y.~Yedelkina,
%``Large-$P_T$ inclusive photoproduction of $J/\psi$ in electron-proton collisions at HERA and the EIC,''
Phys. Lett. B \textbf{811}, 135926 (2020)
doi:10.1016/j.physletb.2020.135926
[arXiv:2009.08264 [hep-ph]].
%19 citations counted in INSPIRE as of 09 Mar 2024

%\cite{Li:2019nlr}
\bibitem{Li:2019nlr}
R.~Li,
%``The production of $J/\psi$ associated with $c\bar{c}$ at ep colliders,''
[arXiv:1912.12822 [hep-ph]].
%0 citations counted in INSPIRE as of 12 Mar 2024

%\cite{Chen:2016hju}
\bibitem{Chen:2016hju}
Z.~Q.~Chen, L.~B.~Chen and C.~F.~Qiao,
%``NLO QCD Corrections for $J/\psi+ c + \bar{c}$ Production in Photon-Photon Collision,''
Phys. Rev. D \textbf{95}, no.3, 036001 (2017)
doi:10.1103/PhysRevD.95.036001
[arXiv:1608.06231 [hep-ph]].
%9 citations counted in INSPIRE as of 12 Mar 2024

%\cite{Yang:2022yxb}
\bibitem{Yang:2022yxb}
H.~Yang, Z.~Q.~Chen and C.~F.~Qiao,
%``NLO QCD corrections to pseudoscalar quarkonium production with two heavy flavors in photon-photon collision,''
Phys. Rev. D \textbf{105}, no.9, 094014 (2022)
doi:10.1103/PhysRevD.105.094014
[arXiv:2203.14204 [hep-ph]].
%2 citations counted in INSPIRE as of 12 Mar 2024

%\cite{Kleiss:1985yh}
\bibitem{Kleiss:1985yh}
R.~Kleiss and W.~J.~Stirling,
%``Spinor Techniques for Calculating p anti-p ---\ensuremath{>} W+- / Z0 + Jets,''
Nucl. Phys. B \textbf{262}, 235-262 (1985)
doi:10.1016/0550-3213(85)90285-8
%755 citations counted in INSPIRE as of 10 Mar 2024

%\cite{Qiao:2003ue}
\bibitem{Qiao:2003ue}
C.~F.~Qiao,
%``A New approach for analytic amplitude calculations,''
Phys. Rev. D \textbf{67}, 097503 (2003)
doi:10.1103/PhysRevD.67.097503
[arXiv:hep-ph/0302128 [hep-ph]].
%7 citations counted in INSPIRE as of 15 Apr 2024

%\cite{Dixon:1996wi}
\bibitem{Dixon:1996wi}
L.~J.~Dixon,
%``Calculating scattering amplitudes efficiently,''
[arXiv:hep-ph/9601359 [hep-ph]].
%755 citations counted in INSPIRE as of 10 Mar 2024

%\cite{Dixon:2013uaa}
\bibitem{Dixon:2013uaa}
L.~J.~Dixon,
%``A brief introduction to modern amplitude methods,''
doi:10.5170/CERN-2014-008.31
[arXiv:1310.5353 [hep-ph]].
%205 citations counted in INSPIRE as of 10 Mar 2024

%\cite{Arkani-Hamed:2017jhn}
\bibitem{Arkani-Hamed:2017jhn}
N.~Arkani-Hamed, T.~C.~Huang and Y.~t.~Huang,
%``Scattering amplitudes for all masses and spins,''
JHEP \textbf{11}, 070 (2021)
doi:10.1007/JHEP11(2021)070
[arXiv:1709.04891 [hep-th]].
%397 citations counted in INSPIRE as of 10 Mar 2024

%\cite{Catani:1996vz}
\bibitem{Catani:1996vz}
S.~Catani and M.~H.~Seymour,
%``A General algorithm for calculating jet cross-sections in NLO QCD,''
Nucl. Phys. B \textbf{485}, 291-419 (1997)
[erratum: Nucl. Phys. B \textbf{510}, 503-504 (1998)]
doi:10.1016/S0550-3213(96)00589-5
[arXiv:hep-ph/9605323 [hep-ph]].
%2198 citations counted in INSPIRE as of 09 Mar 2024

%\cite{Catani:2002hc}
\bibitem{Catani:2002hc}
S.~Catani, S.~Dittmaier, M.~H.~Seymour and Z.~Trocsanyi,
%``The Dipole formalism for next-to-leading order QCD calculations with massive partons,''
Nucl. Phys. B \textbf{627}, 189-265 (2002)
doi:10.1016/S0550-3213(02)00098-6
[arXiv:hep-ph/0201036 [hep-ph]].
%664 citations counted in INSPIRE as of 09 Mar 2024

%\cite{Mastrolia:2012bu}
\bibitem{Mastrolia:2012bu}
P.~Mastrolia, E.~Mirabella and T.~Peraro,
%``Integrand reduction of one-loop scattering amplitudes through Laurent series expansion,''
JHEP \textbf{06}, 095 (2012)
[erratum: JHEP \textbf{11}, 128 (2012)]
doi:10.1007/JHEP11(2012)128
[arXiv:1203.0291 [hep-ph]].
%153 citations counted in INSPIRE as of 25 Apr 2024

%\cite{Peraro:2014cba}
\bibitem{Peraro:2014cba}
T.~Peraro,
%``Ninja: Automated Integrand Reduction via Laurent Expansion for One-Loop Amplitudes,''
Comput. Phys. Commun. \textbf{185}, 2771-2797 (2014)
doi:10.1016/j.cpc.2014.06.017
[arXiv:1403.1229 [hep-ph]].
%141 citations counted in INSPIRE as of 09 Mar 2024

%\cite{Wu:2023upw}
\bibitem{Wu:2023upw}
Z.~Wu, J.~Boehm, R.~Ma, H.~Xu and Y.~Zhang,
%``NeatIBP 1.0, a package generating small-size integration-by-parts relations for Feynman integrals,''
Comput. Phys. Commun. \textbf{295}, 108999 (2024)
doi:10.1016/j.cpc.2023.108999
[arXiv:2305.08783 [hep-ph]].
%14 citations counted in INSPIRE as of 09 Mar 2024

%\cite{Dittmaier:2003bc}
\bibitem{Dittmaier:2003bc}
S.~Dittmaier,
%``Separation of soft and collinear singularities from one loop N point integrals,''
Nucl. Phys. B \textbf{675}, 447-466 (2003)
doi:10.1016/j.nuclphysb.2003.10.003
[arXiv:hep-ph/0308246 [hep-ph]].
%147 citations counted in INSPIRE as of 10 Mar 2024

%\cite{Pumplin:2002vw}
\bibitem{Pumplin:2002vw}
J.~Pumplin, D.~R.~Stump, J.~Huston, H.~L.~Lai, P.~M.~Nadolsky and W.~K.~Tung,
%``New generation of parton distributions with uncertainties from global QCD analysis,''
JHEP \textbf{07}, 012 (2002)
doi:10.1088/1126-6708/2002/07/012
[arXiv:hep-ph/0201195 [hep-ph]].
%6982 citations counted in INSPIRE as of 14 Mar 2024

%\cite{ParticleDataGroup:2020ssz}
\bibitem{ParticleDataGroup:2020ssz}
P.~A.~Zyla \textit{et al.} [Particle Data Group],
%``Review of Particle Physics,''
PTEP \textbf{2020}, no.8, 083C01 (2020)
doi:10.1093/ptep/ptaa104
%3266 citations counted in INSPIRE as of 09 Apr 2022

%\cite{H1:2010udv}
\bibitem{H1:2010udv}
F.~D.~Aaron \textit{et al.} [H1],
%``Inelastic Production of J/psi Mesons in Photoproduction and Deep Inelastic Scattering at HERA,''
Eur. Phys. J. C \textbf{68}, 401-420 (2010)
doi:10.1140/epjc/s10052-010-1376-5
[arXiv:1002.0234 [hep-ex]].
%73 citations counted in INSPIRE as of 10 Mar 2024

%\cite{Li:2009zzu}
\bibitem{Li:2009zzu}
R.~Li and K.~T.~Chao,
%``Photoproduction of $J/psi$ in association with a $c \bar{c}$ pair,''
Phys. Rev. D \textbf{79}, 114020 (2009)
doi:10.1103/PhysRevD.79.114020
[arXiv:0904.1643 [hep-ph]].
%27 citations counted in INSPIRE as of 25 Mar 2024

%\cite{Dulat:2015mca}
\bibitem{Dulat:2015mca}
S.~Dulat, T.~J.~Hou, J.~Gao, M.~Guzzi, J.~Huston, P.~Nadolsky, J.~Pumplin, C.~Schmidt, D.~Stump and C.~P.~Yuan,
%``New parton distribution functions from a global analysis of quantum chromodynamics,''
Phys. Rev. D \textbf{93}, no.3, 033006 (2016)
doi:10.1103/PhysRevD.93.033006
[arXiv:1506.07443 [hep-ph]].
%1856 citations counted in INSPIRE as of 15 Apr 2024

%\cite{Hou:2019qau}
\bibitem{Hou:2019qau}
T.~J.~Hou, K.~Xie, J.~Gao, S.~Dulat, M.~Guzzi, T.~J.~Hobbs, J.~Huston, P.~Nadolsky, J.~Pumplin and C.~Schmidt, \textit{et al.}
%``Progress in the CTEQ-TEA NNLO global QCD analysis,''
[arXiv:1908.11394 [hep-ph]].
%50 citations counted in INSPIRE as of 15 Apr 2024

\end{thebibliography}
\end{document}